# Understanding the Origin of a Second Mobility Reversal in Optoelectrically Powered Metallo-Dielectric Janus Particles

S. S. Das[1], P. García-Sánchez[2], A. Ramos[2], G. Yossifon[1,3]

[1]School of Mechanical Engineering, Tel-Aviv University, Tel-Aviv 69978, Israel

[2]Departamento de Electrónica y Electromagnetismo, Facultad de Física, Universidad de Sevilla, Avenida Reina Mercedes s/n, Sevilla 41012, Spain

[3]Department of Biomedical Engineering, Tel-Aviv University, Tel-Aviv 69978, Israel

**Abstract**

While previous studies indicated the mobility reversal of an electrically-powered metallo-dielectric Janus particle (JP) with increasing frequency, here we report an intriguing second mobility reversal observed in optoelectronically-driven JPs. In contrast to the commonly used setup with parallel ITO-coated glass substrates to induce a uniform electric field orthogonal to the velocity direction, this setup incorporates a thin photoconductive layer deposited on the bottom ITO-coated glass substrate. We have found that the reversal is associated with the asymmetry of the bottom substrate's photoconductivity, localized underneath the JP, resulting from the self-shading effect of the metallic hemisphere under top optical illumination. Numerous control tests, including optical illumination from the bottom, along with numerical simulations, support this hypothesized mechanism.

## 1. Introduction

Active particles that are self-propelled by converting external energy into mechanical motion constitute an exciting and interdisciplinary field of study [1–3]. This research area offers a wide range of potential applications, including drug delivery [4], environmental cleanup and remediation [5], immunosensing [6], microsurgery [7,8], and self-repair systems [9]. The mechanism that propels active particle motion is inherently built into their design, allowing them to asymmetrically draw and dissipate energy, thus generating local force gradients for self-propulsion even in uniform environments [10]. While traditional methods, such as dielectrophoresis [11], diffusiophoresis [12], and magnetophoresis [13], rely on externally imposed gradients to move passive particles, resulting in collective migration in a predetermined direction, self-propelling particles move autonomously along their own individual trajectories [14,15]. This autonomy enables them to operate under simpler field conditions, efficiently navigate confined spaces, exhibit collective behaviors [16,17], and self-assemble into more complex structures [18–21]. Notably, certain particles, like metallodielectric Janus particles (JPs), can self-propel under a uniform AC electric field through induced-charge electrophoresis (ICEP) [22–24] or self-dielectrophoresis (sDEP) [25]. Modulating the frequency of the electric field can trigger different electrokinetic effects, influencing particle locomotion, interactions with other particles (both active and passive), and collective dynamics [26,27].

Despite extensive research over the past decade on the motion dynamics and interaction mechanisms of electrically driven active particles, achieving precise control over the trajectories of these particles - which exhibit autonomous motion and travel along randomly chosen individual paths - remains a significant challenge. However, we have recently discovered [28] that their trajectory can be optoelectronically controlled in a reconfigurable and dynamic manner for each individual active particle. This was attained by generating optically patterned virtual electrodes in an optoelectronic (OE) setup [29–32] comprising a photoconductive substrate made of hydrogenated amorphous silicon (a-Si:H), with the assistance of a digital micromirror device (DMD) and employing metallo-dielectric Janus particles (JPs) as the active particles. Remarkably, the JPs refrained from crossing the boundary of the optical region, thus facilitating the confinement of their motion within a defined area and enabling dynamic shaping of the JP trajectory (Fig.1a). This phenomenon was linked to either the radially inward electrohydrodynamic (EHD) flow [33], originating from the edges of optical pattern and dominant at the low frequency regime (comparable to the RC time of the alternating current electro-convective (ACEO) flow [34], or positive dielectrophoretic attraction force [11] prevailing at the high frequency regime (beyond the RC time of the electric double layer (EDL) induced at the metallic hemisphere of the JP). This technique offers significant operational flexibility in JP guidance by allowing the use of a single applied field to simultaneously propel and precisely steer one or multiple JPs in parallel towards predetermined destinations.

## 2. Materials and Methods
### 2.1. Fabrication of JPs and solution preparation

The stock solution of polystyrene microparticles (Sigma-Aldrich), approximately 27 µm in diameter, was centrifuged and washed three times with isopropanol (IPA). The solution was then pipetted onto a glass slide to form a monolayer of particles as the IPA evaporated. After treatment with oxygen plasma for 1 minute, the slide was sequentially coated with 15 nm of chromium (Cr), 50 nm of nickel (Ni), and 15 nm of gold (Au), following the procedure outlined by Wu et al. 2021 [45]. A similar methodology was used to fabricate Janus particles (JPs) with silica particles (microParticles GmbH), approximately 26.5 µm in diameter. To make polystyrene-ITO JPs (~27 µm in size), the slide was coated with an ~ 200 nm ITO layer. To detach the resulting JPs, the coated glass substrate was sonicated in deionized water (DIW) containing 2% (v/v) Tween 20 (Sigma-Aldrich). The JPs were then rinsed three times with the working medium to minimize their adhesion to the substrate. This working medium was prepared by serially diluting a 1M KCl stock solution to 10 μM KCl and adding 0.1% (v/v) Tween 20, resulting in a conductivity of 5-6 μS/cm.

*2.2. Optoelectronic device fabrication*

The optoelectronic (OE) microfluidic chamber was fabricated by placing a ~120 μm thick spacer layer (Grace-Bio) in between two parallel ~200 nm thick indium tin oxide (ITO)-coated glass slides (Sigma-Aldrich) (see Fig. 1a). The bottom ITO-coated glass slide was further coated with a 900-1000 nm thick photoconductive layer of hydrogenated amorphous silicon (a-Si:H) using plasma-enhanced chemical vapor deposition (PECVD) (Oxford Plasma Instruments, Plasma Lab 100). The a-Si:H layer was generated using the following process recipe: $SiH_4$ (20 sccm), $H_2$ (60 sccm), substrate temperature of 200°C, pressure of 1000 mTorr, and RF power of 30W. To characterize the photoconductivity versus dark conductivity of the fabricated a-Si:H layer, a photoresistor test structure with dimensions of ~ 410 μm × 500 μm was patterned using photolithography techniques (refer Fig. S2a). This structure was fabricated from the a-Si:H layer deposited onto a 1 mm thick regular glass slide, employing the same deposition method as described earlier. Subsequently, the electrical contacts were made by depositing a 50-nm-thick layer of titanium (Ti) followed by a 200 nm thick layer of gold (Au) using an electron-beam evaporator (Fig. S2a). These contacts were then soldered to wires for electrical connectivity.

*2.3. Experimental setup and operation*

The OE system comprises a DMD-based pattern illuminator (Mightex Polygon 1000-G, 625nm, ~1.4W LED source) integrated into an upright microscope (Nikon ECLIPSE Ni-E) equipped with Marzhauser Scan Plus 100 × 85 motorized stages featuring a 4 mm lead screw pitch, a Prime BSI Express Camera, and a Nikon Fluorescence LED source. This setup enables illumination of a light pattern onto the substrate (Fig. 1a). The optical power densities of the light patterns projected through a 10x objective lens (Nikon) were measured at the substrate plane using an Ophir Photonics VEGA power meter. Photoconductivity was estimated by exposing the photoresistor test structure to different optical intensities of 625 nm wavelength light projected through our customized OE system, while simultaneously evaluating its electrical behavior using I-V measurements with a Keithley 2636A source meter. Under a 5V DC voltage and a maximum top illumination intensity of ~0.65 W/cm², the a-Si:H test structure exhibits a more than two orders of magnitude difference between its photoconductivity and dark conductivity (Fig. S2b). For the characterization of JP's motion, both top and bottom illuminations were utilized. Top illumination employed either a 625nm (Mightex LED source) or 621nm (Nikon Fluorescence LED source), while a custom-built bottom illumination setup was designed for bottom or combined top-bottom illumination. This setup includes a 625nm LED source (Mightex) and various optical components (refer to Figure S1a, b). In Figure S1b, a convex lens (Thorlabs, model: LB1757-A) focuses the light beam emitted by the 625nm LED source onto a dichroic mirror (Thorlabs, model: CCM1-G01/M), which then directs the light through a 10x objective lens (Olympus) attached to the bottom plane of the OE substrate, positioned on the microscope stage. Copper tapes (Digi-Key, model: 3M9887-ND) were affixed to both ITO-coated glass slides in the OE device. AC electrical potentials (10-20Vpp, 1kHz-10MHz sine waves) were applied between the two electrode leads connected to the copper tapes using a function generator (Agilent, 33250A) and monitored with an oscilloscope (Tektronix, TPS-2024).

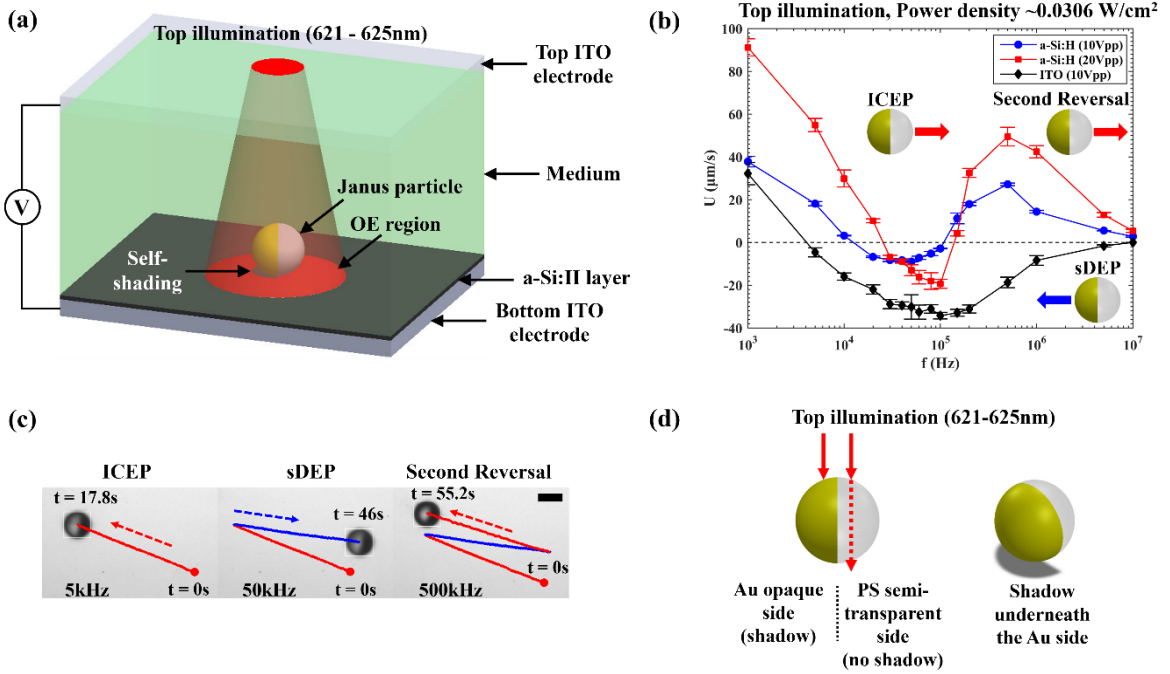

**Fig.1: Experimental evidence of a second mobility reversal and our hypothesized underlying mechanism.** (a) Schematic of the optoelectronic (OE) experimental setup used for studying the mobility of a Janus particle (JP) translating within an optically patterned region illuminated (621-625 nm) from the top, onto a bottom photoconductive substrate (a-Si:H) under an applied electric field. (b) the appearance of a second mobility reversal for JPs moving within the OE setup as delineated in part (a) in contrast to the common two parallel ITO-coated (with additional 15nm $SiO_2$ layer on the bottom ITO-coated electrode) glass slides setup (absent of photoconductive layers) which does not show such second reversal. (c) trajectories of the JPs within the OE setup as delineated in part (a), exhibiting the two mobility reversals depending on the applied electric field frequencies (ICEP: 5kHz, sDEP: 50kHz, second reversal: 500kHz) (scale bar: 30μm). (d) a schematic description of the hypothesized induced symmetry-broken shading underneath the JP's Au-coated hemisphere which in turn results in a symmetry-broken induced photoconductivity of the substrate underneath the JP's dielectric and metallic coated sides.

## 2.4. Microscopy and image analysis

To analyze frequency dispersion, we observed the particle motion in the OE microfluidic chamber using a Nikon ECLIPSE Ni-E microscope equipped with a 10x lens. The observations were recorded using a Prime BSI Express camera for approximately 30-40 seconds at a rate of 5 frames per second. Particle velocities were determined by tracking particle displacement over approximately 150 frames using a custom Python (3.9.7) script developed using the Channel and Spatial Reliability Tracking (CSRT) algorithm from the OpenCV python library. The velocities across mobile particles were then averaged. To validate the accuracy of the Python script, we compared the velocities obtained through this methodology with those derived from manual measurements using ImageJ software.

*2.5. Numerical simulations*

We used COMSOL Multiphysics (v5.3) to determine the electrical force acting on a Janus particle (JP) positioned above an electrode coated with hydrogenated amorphous Silicon (a-Si:H) and illuminated from above. A 3D simulation domain was considered (Fig.2) to solve for the electric potential. The electrical double layers (EDL) at the interfaces were assumed to be negligibly thin, meaning that the EDL thickness at both the particle-electrolyte and a-Si:H-electrolyte interfaces is much smaller than the geometric dimensions of the setup. Under this assumption, the electric potential $\phi(r,t)$ can be obtained from the continuity of the electrical current.

We imposed a ground potential ($\phi = 0$) on the top electrode in Fig.2, while the bottom electrode was subjected to an AC voltage given by $V(t) = V_0 \, Re[exp(i\omega t)]$, where $V_0$ is the voltage amplitude, $\omega$ the angular frequency of the AC voltage, and $Re[exp(i\omega t)]$ represents the real part of the expression in brackets. Thus, the electric potential in the domain can be expressed as $\phi(r,t) = Re[\tilde{\phi}(r) \exp(i\omega t)]$. The current conservation in each medium is then written as $\nabla \cdot [(\sigma + i\omega\varepsilon)\nabla\tilde{\phi}(r)] = 0$, where $\sigma$ and $\varepsilon$ are the electrical conductivity and dielectric permittivity of the medium, respectively. Since the system is illuminated from above, we distinguish between two regions in the a-Si:H layer: the region exposed to light and the region shaded by the metallic side of the Janus particles (see Fig.2). The dielectric part of the JP is (semi-)transparent, so we assume for simplicity that the a-Si:H layer beneath it has the same conductivity as other illuminated areas, while the region shaded by the metallic side of the JP is modeled as an insulator (see Fig.2).

| Electrolyte conductivity | $\sigma_l$ | 0.5 mS/m |
|---|---|---|
| Electrolyte dielectric constant | $\varepsilon_l$ | 80 |
| a-Si:H conductivity (illuminated) | $\sigma_{Si}$ | {0.20, 1.28, 5.12} mS/m |
| a-Si:H dielectric constant | $\varepsilon_{Si}$ | 11.8 |
| Thickness of a-Si:H | $d$ | 1 µm |
| Thickness of Diffuse layer in electrolyte | $\lambda_D$ | 50 nm |
| Diameter of Janus particle | $2R$ | 27 µm |

| Gap between particle and electrode | h | 1% of 2R |
| --- | --- | --- |
| Gap between top and bottom electrodes | H | 120 μm |

**Table 1**: Physical parameters used in numerical simulations.

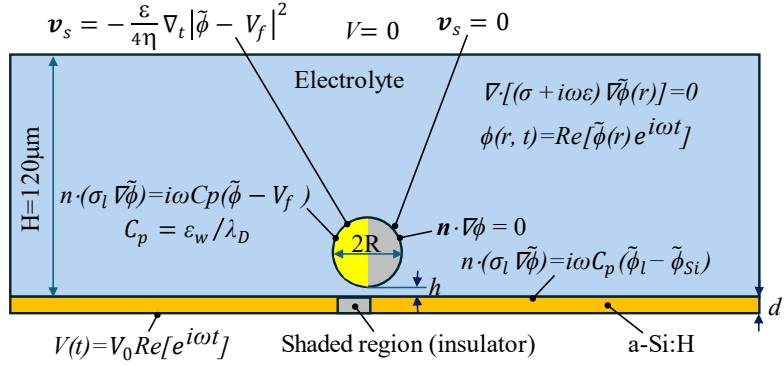

**Fig. 2**: Schematics of the numerical simulation domain geometry (not to scale) and boundary conditions.

The boundary condition imposed on the dielectric side of the JP is that the total current is zero ($n \cdot \nabla \phi = 0$), corresponding to a material with negligible conductivity and a dielectric constant significantly lower than that of the electrolyte. On the metal side of the JP, the electric current induces an EDL at the metal-electrolyte interface. This interface is modeled as a capacitor with capacitance per unit area of $C_p = \varepsilon_w/\lambda_D$, where $\varepsilon_w = 80\varepsilon_0$ is the permittivity of the electrolyte and $\lambda_D$ is the length of the electrolyte Debye, i.e. the thickness of the diffuse ionic layer. Thus, the boundary condition on the metal side of the JP is $n \cdot (\sigma_l \nabla \tilde{\phi}) = i\omega C_p(\tilde{\phi} - V_f)$, where $\sigma_l$ is the conductivity of the electrolyte and $V_f$ is the unknown phasor potential of the JP metal side (a floating conductor). This unknown is determined by imposing the condition of zero total current on the particle surface. Additionally, an EDL is induced at the interface between the electrolyte and illuminated (conducting) Silicon. The boundary condition imposed at this interface is written as $n \cdot (\sigma_l \nabla \tilde{\phi}) = i\omega C_p(\tilde{\phi}_l - \tilde{\phi}_{Si})$, where $(\tilde{\phi}_l - \tilde{\phi}_{Si})$ indicates the voltage drop across the EDL induced at the liquid-silicon interface. Table 1 shows the values of the physical parameters used in the numerical simulations. The total electrical force on the Janus particle is calculated from the solution of the electric potential by integrating the flux of the Maxwell stress tensor through a closed surface that contains the particle.

The particles' velocity due to this force can be estimated using the expression for viscous drag on a sphere moving parallel to a plane. According to Goldman et al. 1967 [46], the viscous drag is given by $F = 6\pi\eta RU[0.9588 - (8/15)\ln(h/R)]$, where $\eta$ is the liquid viscosity, $U$ is the velocity of the particle and $h$ is the distance from the particle to the plane. By equating the viscous drag to the electrical force and applying voltage of 5V between electrodes spaced 120 µm apart, we estimated a maximum velocity of ~700 µm/s for a-Si:H with a conductivity of 5.12 mS/m and a sphere diameter of $2R= 27$ µm. The electrohydrodynamic velocity of the particle is driven by Induced-charge Electroosmosis (ICEO) on the metal side of the particle [22]. ICEO flow is caused by the interaction of the applied electric field with the EDL of the metal particle, resulting in a nonzero time-averaged slip velocity, expressed as:

$$v_s = -\frac{\varepsilon}{4\eta} \nabla_t \left|\tilde{\phi} - V_f\right|^2, \quad (1)$$

where $\eta$ is the electrolyte viscosity, $\nabla_t$ is the tangential component of the gradient operator along the particle surface, and $V_f$ is the amplitude of the electric potential of the floating metal particle. We calculated the fluid velocity field around a static particle by solving the Stokes equation with boundary condition (1) on the metal side of the JP, and no-slip condition on all other boundaries. The ICEP force acting on the particle was determined by integrating the hydrodynamic stresses.

## 3. Results and Discussion

### 3.1. The appearance of a second mobility reversal

Interestingly, while the reversal of mobility direction from ICEP to sDEP with increasing field frequency was previously reported [25] and recently analyzed [35], herein, we have encountered a surprising second mobility reversal from sDEP to a different mode of electrokinetic propulsion with further increase of the frequency (see Fig.1b, c). Similar to earlier findings, here, the JP is again noted to move with its dielectric side forward in ICEP mode within the lower frequency regime (characteristic of the RC time of the induced EDL associated with the JP's metallic coating), transitioning its motion direction to the metallic side forward in sDEP mode with a further increase in frequency. Surprisingly, upon reaching the higher frequency regime (>100kHz), the JP was observed to revert to its initial motion direction, moving with its dielectric side forward, akin to the ICEP mode of propulsion (see Fig.1b-c, and Supplementary Video 1). We refer to this new propulsion mechanism as self-shading optoelectronic modulated electrokinetic propulsion (ss-OMEP). This extends our earlier OMEP phenomenon [36] by incorporating the additional effect of asymmetric shading.

Such a second mobility reversal was not observed in the standard parallel ITO coated glass slides apparatus (see Fig.1b, and Supplementary Video 2). This has suggested that the second mobility reversal effect must be related to the experimental setup of the optoelectronic actuation. A hypothesis that we made and investigated herein, was that the possible asymmetric shading of the photoconductive substrate immediately underneath the JP when illuminating from top is responsible for the second reversal. As schematically depicted in Fig. 1d, the optical transmission of red light (621-625 nm) through the metallic-coated hemisphere of the JP, which is inherently opaque [37] compared to the bare and semi-transparent polystyrene hemispheres [38], results in a contrast in transmitted illumination intensity. This contrast interacts with the photoconductive substrate beneath the JP, causing more shading under the metallic-coated side than the polystyrene side. Consequently, this affects the photoconductivity of the substrate underneath the JP, inducing locally non-uniform photoconductivity.

## 3.2. The effect of combined top and bottom illumination

To further investigate this hypothesis, we configured our optoelectronic apparatus to include an additional illumination source from below, operating concurrently with the top illumination (see Fig. 3a, and Fig. S1). When using only bottom illumination, no second mobility reversal was observed (see Fig. 3b, and Supplementary Video 3). However, when combining weak bottom illumination with strong top illumination or using solely top illumination, we clearly observed the second reversal (see Fig. 3b). Furthermore, under conditions of combined top and bottom illumination, the mobility at the second reversal mode decreased compared to using only top illumination. This decrease is attributed to the reduction in asymmetric shading effects caused by weak bottom illumination. Upon further increasing the bottom illumination intensity, the second reversal disappeared in the case of strong bottom and weak top illumination overlap (see Fig. 3c, and Supplementary Video 4). Here, the JP movement was observed solely in sDEP mode across the entire higher frequency regime. This disappearance correlates with the saturation of the bottom substrate's photoconductivity (see Fig. S2), which suppresses the asymmetric photoconductivity originally induced by the shadowing effect of the JP's metallic hemisphere under top illumination. Additional experimental measurements are presented in Supporting Figures S3 and S4 for a different batch of JPs, with similar observations. The data in Fig. S4a illustrates how altering the intensity of bottom illumination, while maintaining a constant top illumination intensity, gradually eliminates the asymmetric shading effect and establishes a definitive threshold intensity limit. On the other hand, the variation from weak to strong bottom illumination intensity in Fig. S4b does not lead to the emergence of a second mobility reversal phenomenon. Furthermore, we identified a critical threshold in the photoconductive layer thickness necessary for the occurrence of the second mobility reversal phenomenon (see Fig. S5), where the transition frequency from sDEP to the second mobility reversal is found to be significantly dependent on the thickness of a-Si:H.

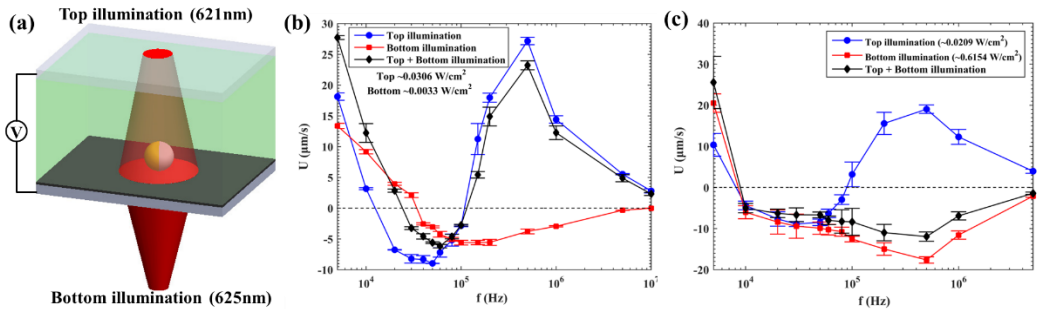

**Fig.3: Investigating the effect of inducing a shadow versus the absence of shadow under top and bottom illuminations, respectively**. (a) Schematics of the experimental setup involving independently controlled top (621 nm) and bottom (625 nm) illuminations. (b) experimentally measured JP velocity versus frequency for either top, bottom or a combined top and bottom illuminations, indicting the absence of second mobility reversal under bottom illumination only, with no shadow effect present. (c) additional experimental results for bottom illumination with higher intensity under the conditions of saturated photoconductivity of the bottom a-Si:H substrate that overwhelms the shadow effect stemming from the top illumination.

*3.3. Numerical simulations and additional control experiments*

The experimental observations in Fig. 3 were complemented by three-dimensional numerical simulations (details provided in the Supporting Materials). The main results, shown in Fig. 4, present the calculated dimensionless horizontal component of the forces (both electrostatic (see Fig. 4a) and electro-hydrodynamic (see Fig. 4b)) acting on the Janus particle (JP) as a function of the dimensionless angular frequency $\Omega$ for varying a-Si:H conductivity, while accounting for electric double layers induced on both the metallic-coated hemisphere of the JP and within the electrolyte interfacing the photoconducting substrate. Positive force values indicate movement towards the dielectric side of the Janus particle.

Interestingly, the electrostatic force (see Fig. 4a), obtained via the integration of the traction force associated with the Maxwell stress tensor, not only exhibits the expected sDEP but also reveals a pronounced reversed force associated with the second mobility reversal of the JP. In contrast, when shading is absent beneath the metallic-coated hemisphere of the JP, the second reversal force disappears, leaving only the sDEP contribution (see, for example, the comparison between the continuous and dashed red lines for the 1.28 mS/m a-Si:H substrate conductivity, corresponding to the cases with and without shading, respectively).

The electrohydrodynamic force, shown in Fig. 4b, is always positive, indicating that the JP moves with its dielectric hemisphere forward, as expected due to the ICEP contribution, regardless of the shading effect. The total force depicted in Fig. 4c is a superposition of the electrostatic force and a scaled-down electrohydrodynamic force (by a factor of 0.05), which is commonly assumed due to the hindrance of the induced zeta-potential by the Stern layer [33,39]. As shown, the simulations predict the two reversals in particle motion (see, for example, the blue continuous line corresponding to 0.2 mS/m a-Si:H layer's conductivity), in qualitative agreement with the experiments when considering the self-shading effect. However, in the absence of shading, the second reversal nearly disappears, leaving only the expected single ICEP-to-sDEP mobility reversal.

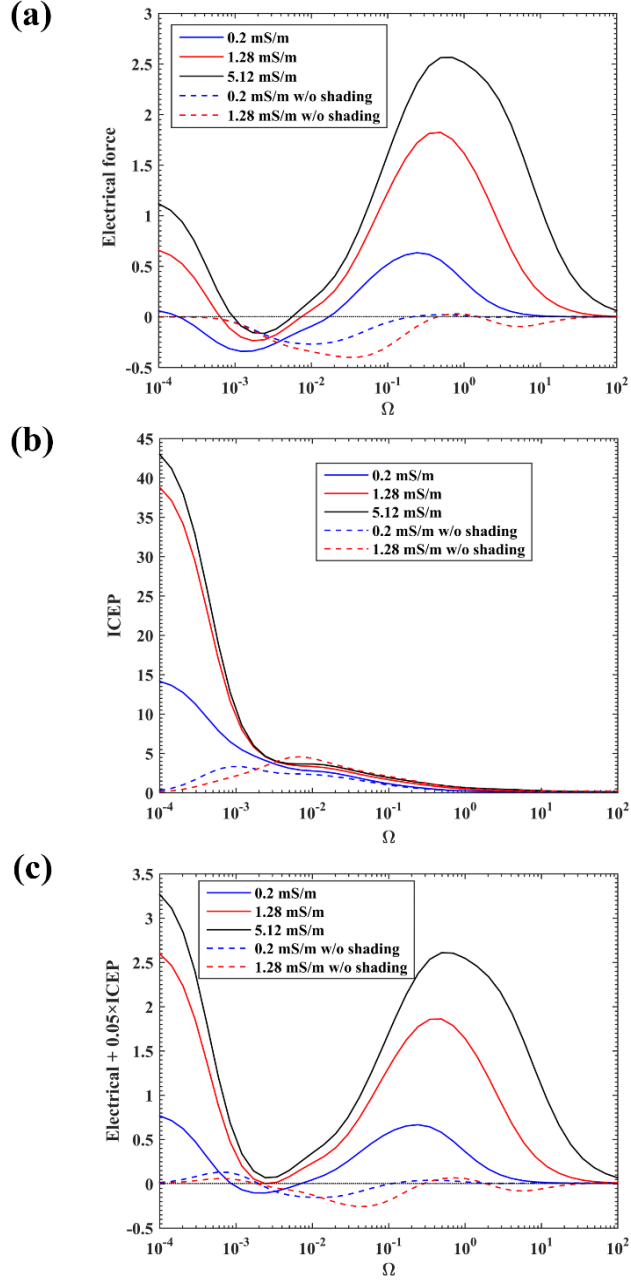

**Fig.4: Three-dimensional numerical simulations of the forces acting on the Janus particle (JP) as a function of frequency:** (a) electrostatic force, (b) electro-hydrodynamic force leading to induced-charge electro-phoresis (ICEP) motion, and (c) total force, which is the superposition of the electrostatic and electro-hydrodynamic forces, with the hydrodynamic force scaled by a factor of 0.05. The forces are calculated for varying photoconductive substrate (a-Si:H) conductivity, both with and without shading (i.e., the region underneath the JP's metallic-coated hemisphere is assumed to be insulating). In these figures the force and frequency were normalized by $\varepsilon_l R^2 E_0^2$ and $\sigma_l/\varepsilon_l$ (inverse of the electrolyte charge relaxation time), respectively, where $E_0$ represents the amplitude of the applied electric field, $\varepsilon_l$ and $\sigma_l$ are the electrolyte's permittivity and conductivity, respectively, and $R$ is the JP's radius.

To further support our hypothesis, we conducted additional experiments using SiO$_2$-Au JPs featuring a silicon dioxide (SiO$_2$) dielectric hemisphere with enhanced optical transparency [40] compared to the polystyrene (PS) dielectric hemisphere [41] in PS-Au JPs. The enhanced transparency of the SiO$_2$ hemisphere is expected to induce a more pronounced asymmetric photoconductivity beneath the JP compared to the PS hemisphere. As shown in Fig. S6a, we observed a second mobility reversal under top illumination, along with its disappearance upon strong bottom illumination in Fig. S6b, attributed to the saturation of photoconductivity in the bottom substrate. As a control experiment, we employed a JP with minimal contrast in optical properties between its hemispheres. This JP consisted of a semi-transparent polystyrene (PS) sphere half-coated with a 200 nm thick layer of fully transparent ITO. This configuration exhibited negligible self-shading effects and consequently minimal asymmetry in the photoconductivity of the bottom substrate beneath the JP, leading to the absence of the second mobility reversal (see Fig. S7).

*3.4. The effect of varying the top illumination power density on the JP mobility*

To delve deeper into the underlying physics, Fig. 5a (see also the Supplementary Videos 4-7) illustrates the impact of varying top illumination intensity on JP mobility. It was observed that at low illumination intensities, the JP exhibits all three modes of motion, including two mobility reversals: one from ICEP to sDEP, and another from sDEP to ss-OMEP mode. However, with increasing intensity, the sDEP mode at intermediate frequencies diminishes, resulting in JP motion exclusively in ICEP or ss-OMEP mode across the entire frequency range, accompanied by a monotonic increase in mobility absolute value. This trend is more clearly depicted in Fig. 5b, which derives from the data in Fig. 5a, showing JP mobility as a function of illumination power density. When compared to the numerical simulations depicted in Fig. 4 for different substrate photoconductivity values outside the shaded region beneath the JP's metallic-coated hemisphere - corresponding to varying top illumination intensities - we observe that higher substrate conductivity (i.e., at higher illumination intensity) enhances the magnitude of the ss-OMEP mobility, which qualitatively aligns with the experimental observations shown in Fig. 5.

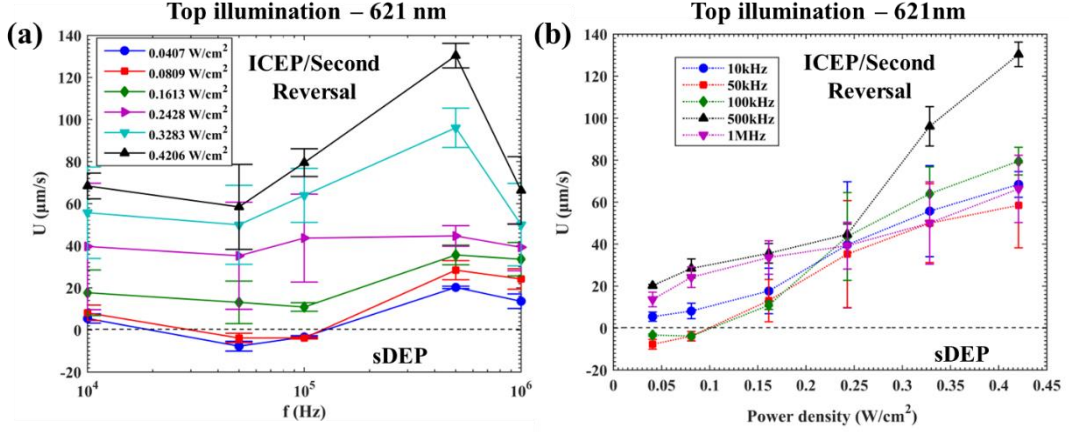

**Fig.5: Investigating the effect of varying the top illumination power density on the JP mobility and second reversal.** (a) Experimentally measured JP velocity versus frequency for varying top illumination power density. (b) Plotting the JP velocity versus the power density obtained from part (a), indicating the disappearance of sDEP mode (JP moving with its metallic side forward) within the intermediate frequency regime, with increasing illumination power density.

Interestingly, the simulations also show a significant increase in velocity magnitude with higher illumination power density in the low-frequency regime, corresponding to the dominant ICEP mode, consistent with the experimental results shown in Fig. 5. As a result of the increased ICEP and ss-OMEP velocities, the sDEP regime vanishes above a certain illumination intensity threshold. To rule out thermophoresis as a potential mechanism - where the metallic side of the JP could be heated upon illumination and driven thermophoretically [42,43] due to induced temperature gradients - we conducted experiments using a setup of two parallel ITO-coated glass slides (with an additional 15 nm $SiO_2$ layer on the bottom ITO-coated electrode). In this setup, particles were illuminated without an electric field, while the metallo-dielectric interface of the JP was aligned using two static magnets placed opposite to each other. We also applied a high-frequency electric field to electro-orient the metallo-dielectric interface without inducing electrokinetic mobility.

We observed no movement in either case. Although thermophoresis appears negligible, we cannot rule out the possibility of electrothermal effects [44], which could contribute to propulsion due to the interaction between locally induced temperature gradients and the applied electric field. However, since our numerical simulations qualitatively match the experimental data, these effects seem to have minimal or no significance.

## 4. Conclusion

To conclude, we have studied a surprising second mobility reversal phenomenon, which we attribute to the self-shading of the JP under top optical illumination. This effect breaks the symmetry of the bottom substrate's photoconductivity directly beneath the JP. We verified this experimentally by varying the ratios of top to bottom illumination intensities and by using particles with different optical transmission properties between their two hemispheres. Additionally, we explored the effect of illumination intensity, demonstrating the disappearance of the sDEP motion mode. Beyond a specific threshold intensity, JP motion was limited to either ICEP or the newly discovered ss-OMEP motion mode under top illumination. These experimental findings were further supported by numerical simulations. Beyond its fundamental significance and relevance for controlling JPs in optoelectronic systems—enabling reprogrammable and parallelized trajectory control of multiple JPs and their directed self-assembly [28]—this self-shading mechanism enhances the optical modulation capabilities of electrically powered JPs. The asymmetric optical transparency of the two JP hemispheres, along with the direction of illumination (from above to induce shading or from below to suppress it), serves as an additional control parameter for modulating electrokinetic mobility and distinguishing between different sub-populations of JPs based on their asymmetric optical and electrical properties.

**Acknowledgments**

G.Y. acknowledges support from the Israel Science Foundation (ISF) (1934/20). P.G.S. and A.R. acknowledge the financial support from MCIN/AEI/10.13039/501100011033/FEDER, UE (Grant No. PID2022-138890NB-I00). We thank the Center for Nanoscience and Nanotechnology, Hebrew University of Jerusalem for assisting us in fabricating the photoconductive substrates.

**Supporting information**

# Understanding the Origin of a Second Mobility Reversal in Optoelectrically Powered Metallo-Dielectric Janus Particles

S. S. Das[1], P. García-Sánchez[2], A. Ramos[2], G. Yossifon[1]

[1]School of Mechanical Engineering, Tel-Aviv University, Tel-Aviv 69978, Israel

[2]Departamento de Electrónica y Electromagnetismo, Facultad de Física, Universidad de Sevilla, Avenida Reina Mercedes s/n, Sevilla 41012, Spain

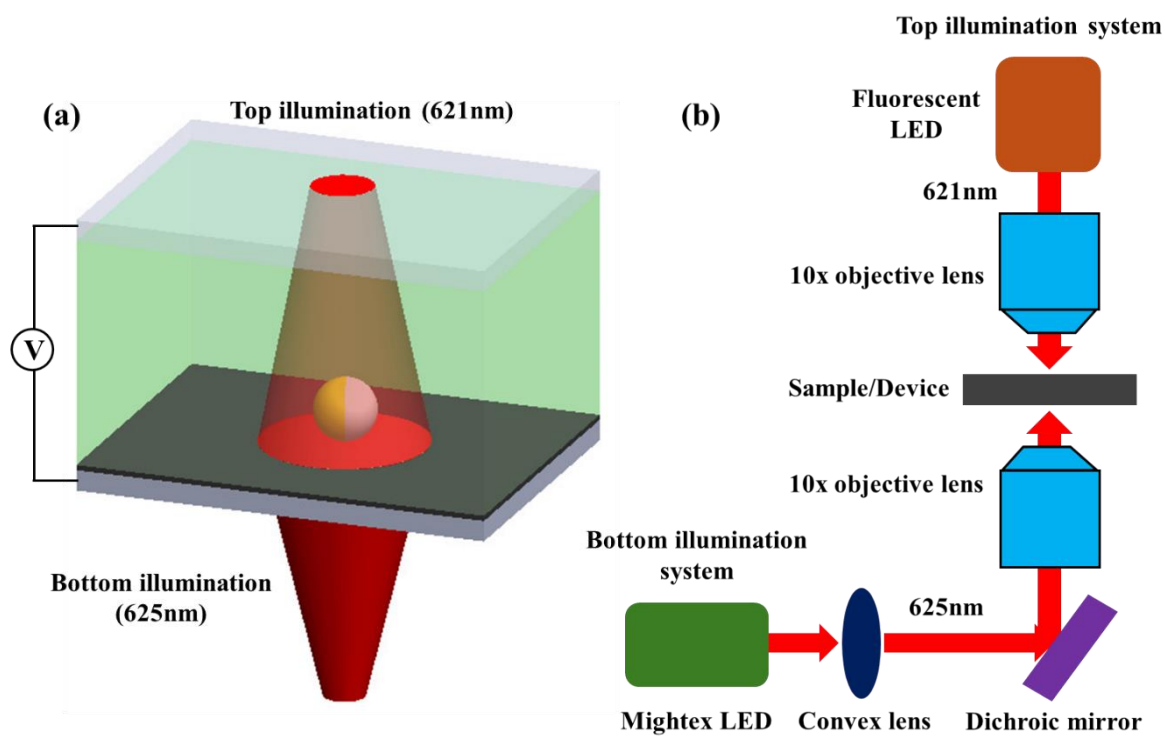

**FIG. S1. Experimental setup.** (a) Schematics of the experimental setup and the corresponding (b) optical setup for simultaneous top and bottom illumination.

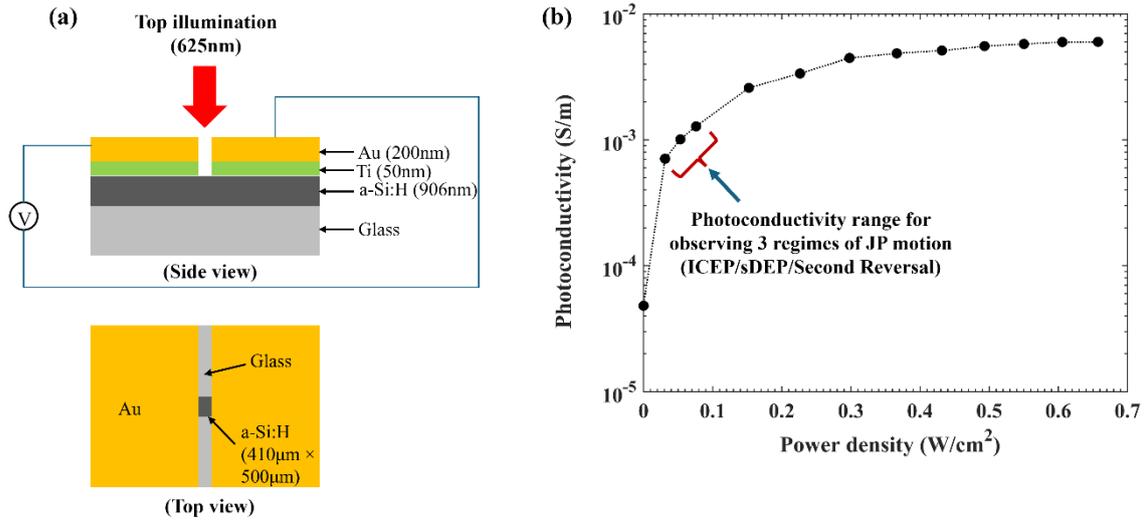

**FIG. S2. Photoconductivity characterization**. (a) Structure of the fabricated test structure used for characterization of the photoconductivity a-Si:H layer. (b) Measured photoconductivity versus power density.

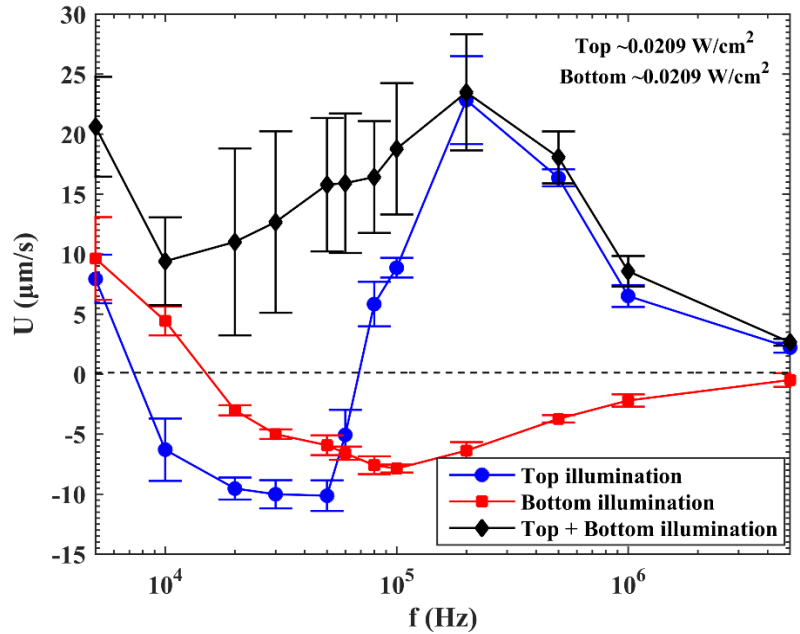

**FIG. S3.** Experimentally measured JP velocity versus frequency for either top, bottom or a combined top and bottom illuminations, of comparable illumination power density.

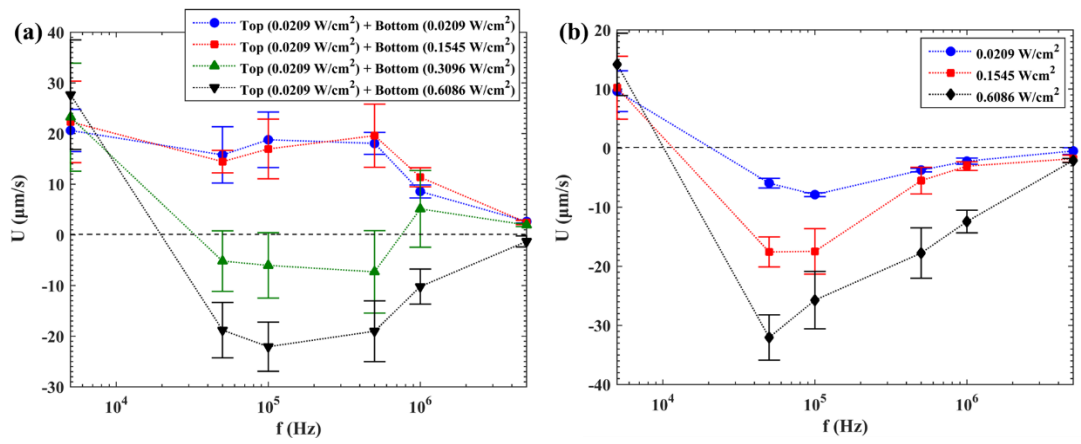

**FIG. S4.** Experimentally measured JP velocity versus frequency for: (a) fixed top and varying bottom illumination power density, (b) only varying bottom illumination power density.

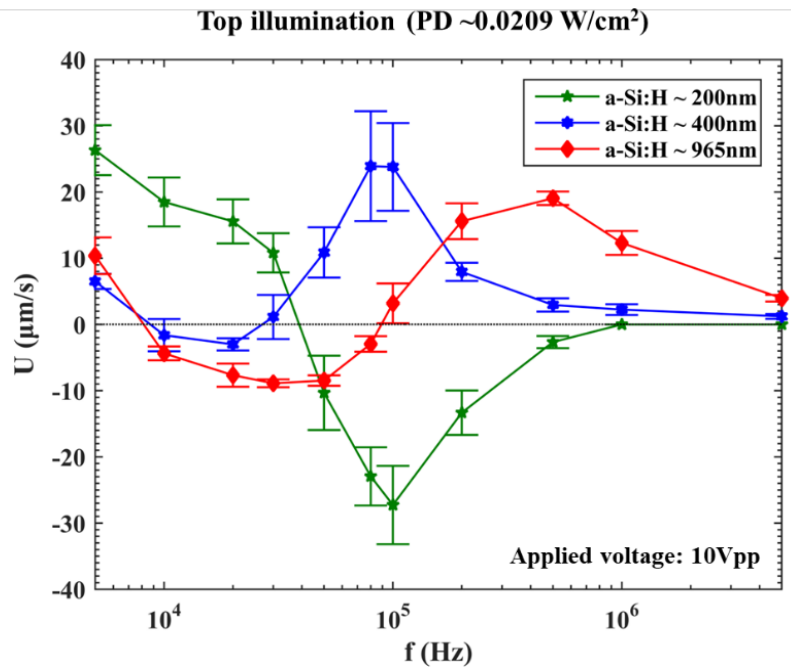

**FIG. S5.** Motion characteristics of polystyrene-gold (PS-Au) Janus particle (JP) (diameter of 27μm) under top illumination in optoelectronic (OE) device made of varying a-Si:H layer thicknesses.

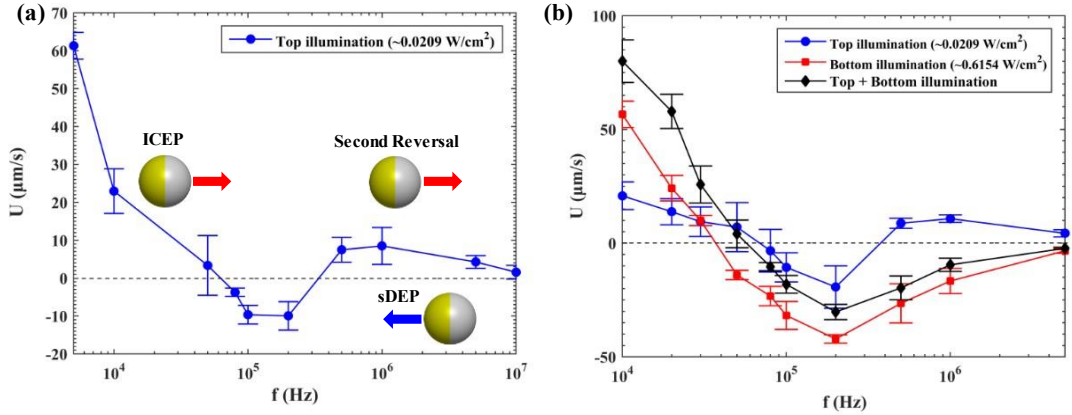

**FIG. S6. Existence of a second mobility reversal is also observed in the silicon dioxide (SiO$_2$) dielectric side, which has enhanced optical transparency compared to the polystyrene (PS) hemisphere.** (a) Experimentally measured JP velocity versus frequency for top illumination indicating a response of the Au coated SiO$_2$ JPs (diameter: 26.5 μm) that is in qualitative agreement to that of the Au coated PS JPs (diameter: 27μm). (b) Another experimental evidence of the disappearance of the second mobility reversal under conditions of saturation of bottom a-Si:H substrate's photoconductivity upon intense bottom illumination.

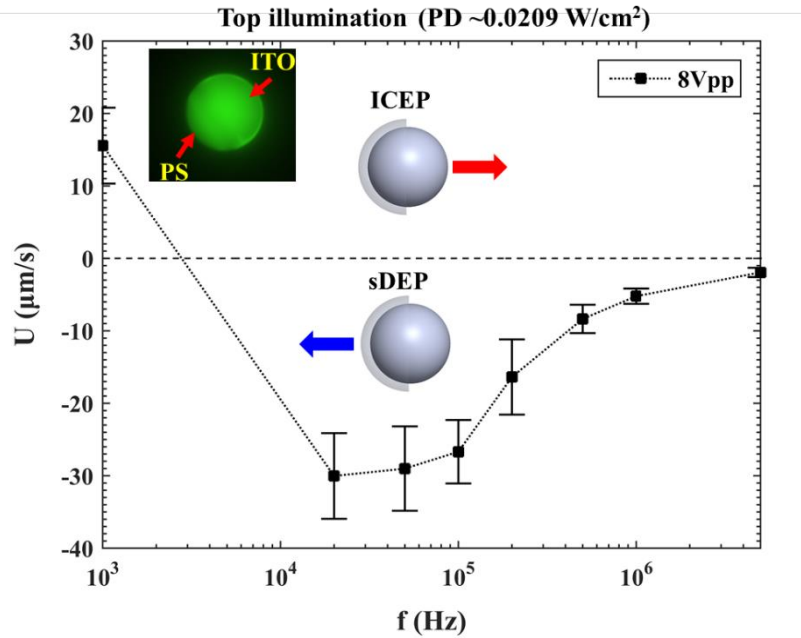

**FIG. S7. Control experiment using a transparent JP made of polystyrene (PS) sphere half-coated with ITO, without presence of any shadow effect.** Experimentally measured JP velocity versus frequency for top illumination indicating a response of an ITO (200 nm) coated PS JP indicating the absence of a second mobility reversal.

**Supplementary Videos:**

Supporting Information is available from the Online Library or from the author.

**Supplementary Video 1:** Electrokinetic propulsion of a Janus particle within an opto-electronic system under top illumination demonstrates two instances of mobility reversal.

**Supplementary Video 2:** Electrokinetic propulsion of a Janus particle within conventional parallel ITO-coated glass slides demonstrates a single mobility reversal.

**Supplementary Video 3:** Electrokinetic propulsion of a Janus particle within an opto-electronic system under bottom illumination demonstrates a single mobility reversal.

**Supplementary Video 4:** The disappearance of the second mobility reversal of Janus particles in an opto-electronic device under combined top and strong bottom illumination.

**Supplementary Video 5:** The effect of varying top illumination intensity on the electrokinetic propulsion of a Janus particle at a 10 kHz electric field frequency.

**Supplementary Video 6:** The effect of varying top illumination intensity on the electrokinetic propulsion of a Janus particle at a 50 kHz electric field frequency.

**Supplementary Video 7:** The effect of varying top illumination intensity on the electrokinetic propulsion of a Janus particle at a 500 kHz electric field frequency.